# Scanning Tunneling Spectroscopy of MoS$_2$ monolayer in presence of ethanol gas

*Seyed Ali Hosseini[†], Azam Iraji zad [†, ‡], Masud Berahman[†], Farzaneh Aghakhani Mahyari[‡], Seyed Hossein Hosseini Shokouh[†]*

† Department of physics, Sharif university of technology, Tehran, Iran

‡ Institute for nanoscience and nanotechnology, Sharif university of technology, Tehran, Iran

**Corresponding Author:**    Azam Irajizad (**iraji@sharif.edu**)

**Abstract**

Due to high surface to volume ratio and tunable band gap, 2 Dimensional (2D) layered materials such as MoS$_2$, is good candidate for gas sensing applications. This research mainly focuses on variation of Density of States (DOS) of MoS$_2$ nanosheets caused by ethanol adsorption. The nanosheets are synthesized by liquid exfoliation, and then using Scanning Tunneling Spectroscopy (STS) and Density Functional Theory (DFT), local electronic characteristic such as DOS and band gap in non-vacuum condition are analyzed. The results show that ethanol adsorption enhances DOS and deform orbitals near the valence and conduction bands that increase transport of carriers on the sheet.

**Introduction**

Transition Metal Dichalcogenides (TMD), as a new family of 2D materials with the chemical formula $MX_2$, have attracted much attention [1]. These materials are composed of a transition metal (M) of group 4, 5 or 6 with Sulfur, Selenide or Telluride (X) which show wide electrical characteristic including conductors, semiconductors and superconductors [2-6]. These materials have a layered structure held together with van der Waals forces. $MoS_2$, the most explored TMD, has a tunable energy gap and presents many applications in field of catalysts, solar energy, electronics, optoelectronics and gas sensing [7-11].

During the past decades, physical properties of the bulk $MoS_2$ with indirect energy gap and layered structure has been investigated extensively [12-16]. Hexagonal and octahedral are the most common crystalline structures for $MoS_2$ with semiconductor and metallic properties, respectively [17]. Mono and few layer $MoS_2$ can be easily exfoliated from the bulk due to layered structure by mechanical exfoliation method [18,19].

It is known that electronic properties including energy gap of $MoS_2$ are strongly related to the number of atomic layers. In bulk state, $MoS_2$ has an indirect energy gap around 1.29eV whereas reducing the number of layers, due to quantum confinement, not only increases the optical energy gap up to 1.9eV but also makes it a direct gap semiconductor [18,20,21]. Such transition from indirect to direct gap promotes many applications in the electronics and optoelectronics industries [22-29]. Due to high exciton binding energy, the quasi-particle band gap is 2.17eV while the optical one is about 1.9eV [30]. In addition, the optical energy gap can be strongly affected by dielectric property of the environment and can vary from 1.9eV to 2.8eV [31]. For example, it is observed that the TMD sandwich between the graphene or boron nitride changed the exciton binding energy and energy gap [32]. Moreover, large surface to volume ratio of few layers' system provides a high number of interaction sites to gas environment resulting in variation of electrical properties. $MoS_2$ based field effect transistors have recently been investigated for sensing of $NO_2$, NO, $NH_3$ [33-35]. Other investigations indicate that non-polar gases like chloroform and toluene that have low dielectric coefficient cause lower current and polar gases like ethanol, methanol and acetonitrile with higher dielectric coefficient induce charges on $MoS_2$ and cause higher current [36]. Although there are various reports on gas sensing characteristics of $MoS_2$, but the mechanism behind the interaction of gas molecules and $MoS_2$ surface is ambiguous [36].

STS can measured electronic properties of materials such as DOS and energy band gap. By using STS in various environmental conditions, effect of this conditions investigated and useful information on the operation of electronic devices including gas sensors is achieved.

This research is mainly focused on electronic DOS variations due to ethanol adsorption on $MoS_2$ surface using STS method. Our results have shown that ethanol molecules increase the DOS in edges of valance and conduction bands. Additionally, the STS results have been described by DFT simulation using Siesta software. The results show that ethanol adsorption enhances DOS and deform the orbitals near the valence and conduction bands that cause increase in carriers' transport.

**Method**

$MoS_2$ nanosheets were synthesized using liquid exfoliation under sonication process. 100 mg $MoS_2$ powder (from Sigma Aldrich average particle size ~6 μm) was dispersed in 20 mL N-Methyl-2-pyrrolidone (NMP) as liquid dispersant and was sonicated at 80% power cycle for one hour. The resulting suspension was centrifuged for 45min at 1500 rpm. After extraction the thinner and lighter sheets, they were dispersed in Isopropyl Alcohol (IPA) for analysis. Structural study was performed using Scanning Electron Microscopy (SEM), Atomic Force Microscopy (AFM), UV-Vis and Fourier transform infrared spectroscopy (FTIR) spectroscopes.

Scanning Tunneling Microscope (model NAMA SS2) was used to investigate DOS of the samples in presence and absence of ethanol gas. For STS measurements, a droplet of $MoS_2$ in IPA was poured over the surface of a refreshed HOPG substrate under the testing gas environment.

In addition to experimental study, simulation was performed based on density functional theory on the system. The Local-density approximations (LDA) with Perdew-Zunger (PZ) parameterization have been used to investigate the interactions in both the pristine $MoS_2$ and in presence of ethanol [37]. All calculations have been performed using Siesta Package [38].

**Result & Discussion**

UV-visible transmission spectra of the dispersed sheets in IPA in Figure 1(a) indicates two separated exciton peaks at about 625 nm and 685 nm. Raman spectra of the deposited sheets on $SiO_2$ substrate shows a strong in-plane vibrational mode for the Mo and S atoms ($E_{2g}$) and an out-of-plane vibrational mode for the S atoms ($A_{1g}$) (Figure 1(b)). Both data are the characteristic spectra for $MoS_2$.

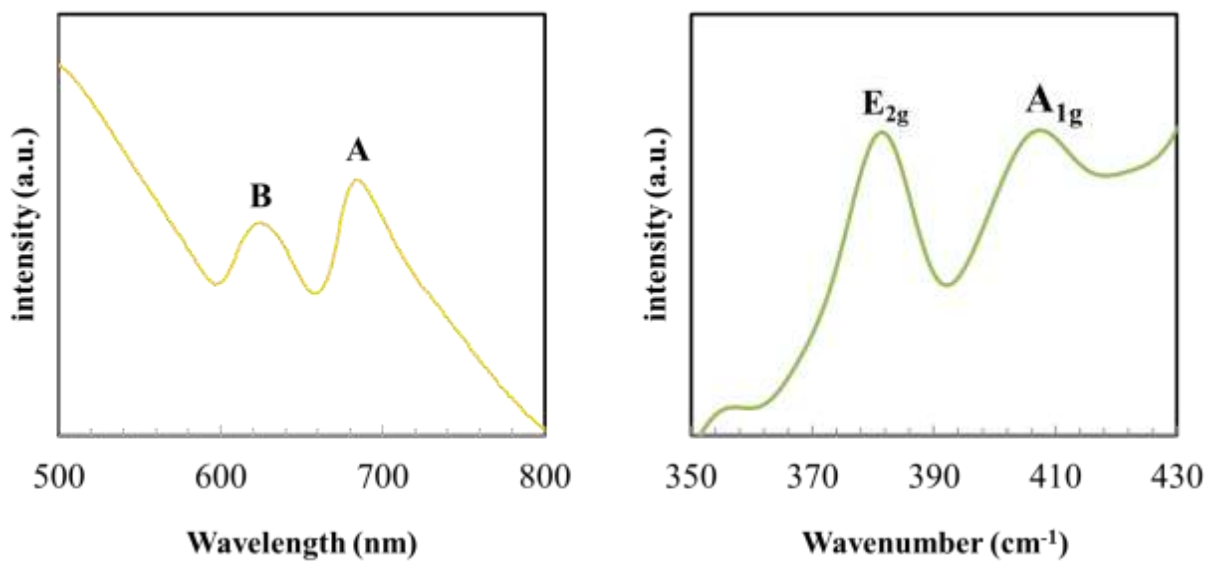

**Figure 1.** a) UV-Vis spectrum of the $MoS_2$ sheets in IPA solvent. Two exciton peak of $MoS_2$ are observed. b) Raman spectrum of the $MoS_2$ sheets shows two characteristic peaks ($E_{2g}$ and $A_{1g}$) of $MoS_2$ sheets.

Microscopic observations of the exfoliated sheets indicate that our preparation method resulted in nanosheets with size around 50 nm up to 1 µm and thickness of mono to few layers as is shown in Figure2 (a) and (b).

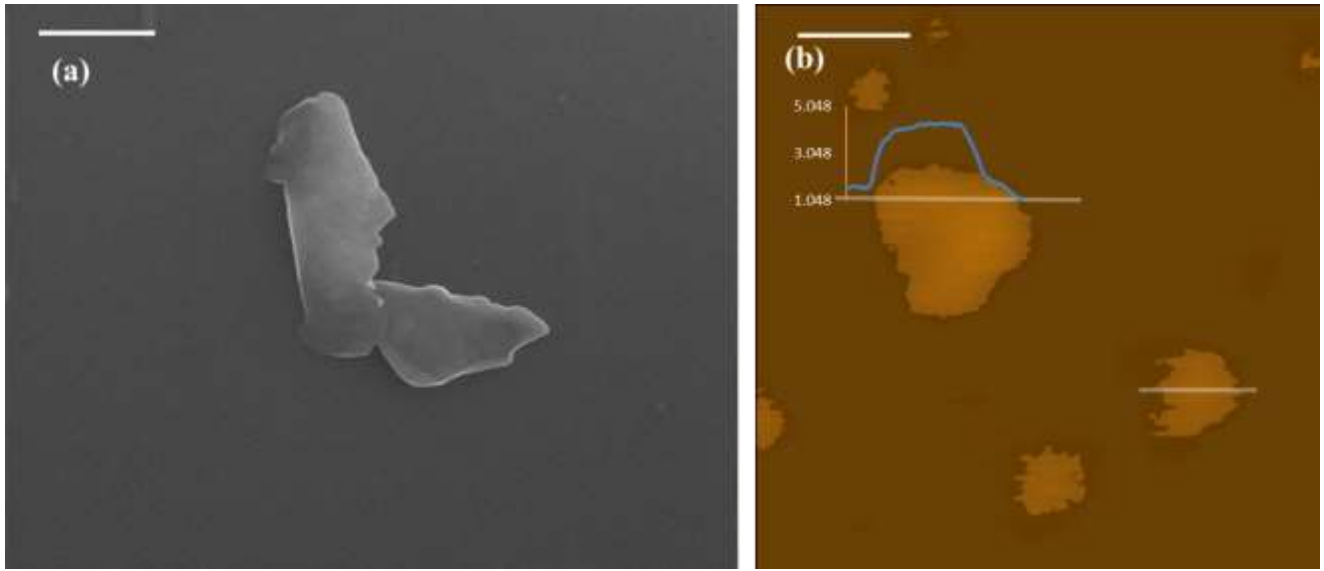

**Figure 2.** a) A typical SEM images and b) AFM images of few layers of MoS$_2$ nanosheet on SiO$_2$ substrate. Scale bar for (a) and (b) are 200 nm and 100 nm respectively.

To study detail of ethanol molecules interaction with the nanosheets, FTIR analysis was applied in absence and presence of ethanol gas. Figure 3 indicates presence of one rather strong peak around 1100 cm$^{-1}$ related to physisorbed ethanol molecules [39]. Therefore, we expect ethanol sensing potential for MoS$_2$ nanosheets.

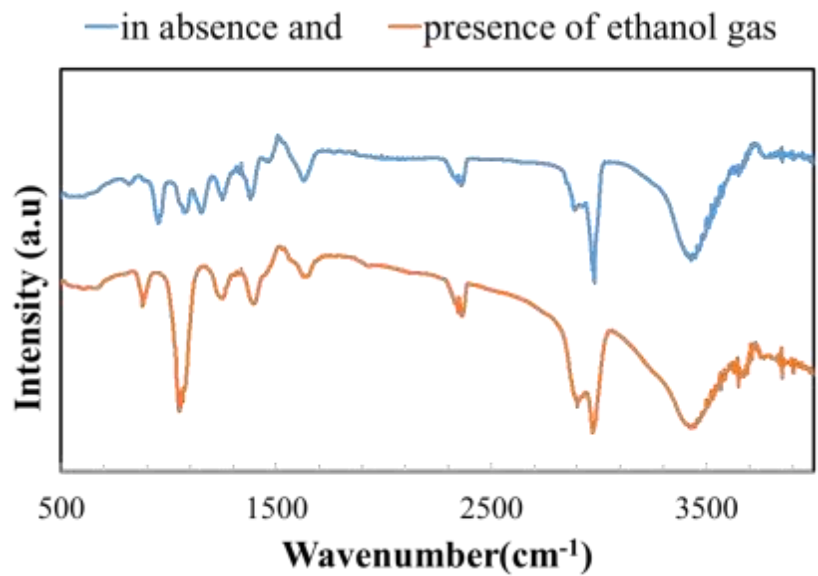

**Figure3:** FTIR spectra of the MoS$_2$ nanosheets in absence (blue) and presence (red) of ethanol gas.

To understand the effect of ethanol on electronic properties of MoS$_2$ surface, we measured DOS variation using STS setup as is shown in Figure 4(a). About 2.15 eV quasi-particle band gap was measured which is in good agreement with the previous reports (Figure 4(b)) [30]. After gas introduction, we observed slightly band gap change and a notable DOS enhancement around band edges compared with pristine MoS$_2$. As the inset image in Figure 4(b) indicates, ethanol adsorption has moved conduction band away about 0.06 eV. Therefore, we expect reduction in charge carriers i.e. decrease in electrical conductivity while in previous studies an increase in sample conductance are reported [36].

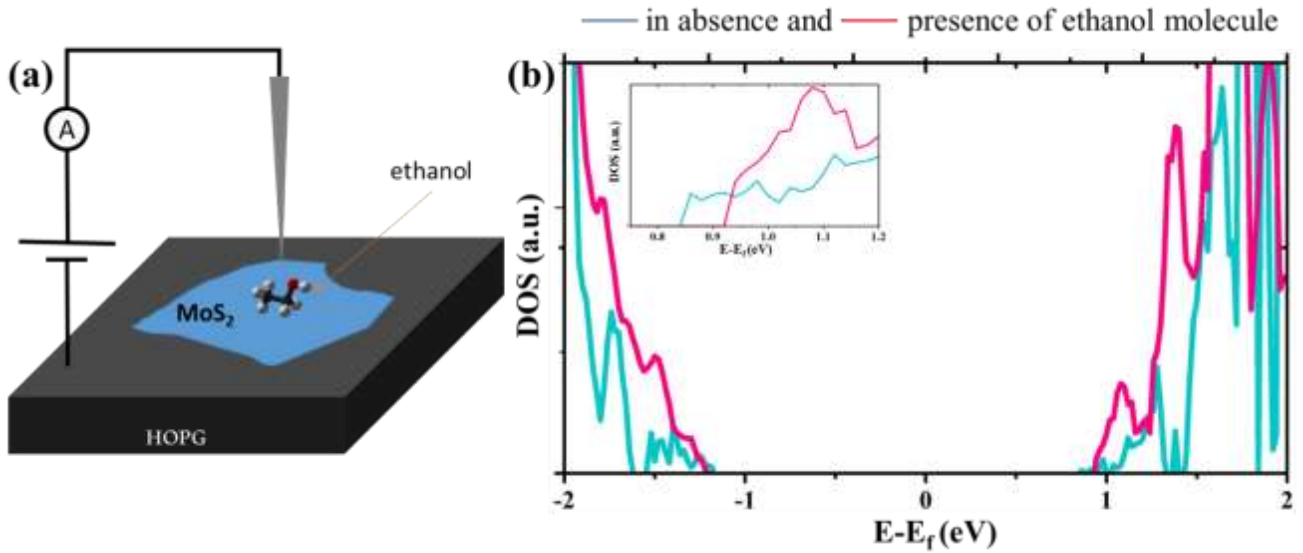

**Figure 4:** a) Schematic of the experimental setup. b) STS data of a typical MoS$_2$ sample in presence (red) and absence (blue) of ethanol gas.

To study the effect of DOS variation on carriers' transport on MoS$_2$ monolayer, DFT calculation was performed. The optimized structure in presence and absence of ethanol shows that ethanol molecules adsorb on the MoS₂ surface by van der Waals interaction which is in good agreement with our FTIR results. Figure 5 exhibits DOS and band structure of the optimized MoS₂ in presence and absence of ethanol that are in good agreement with the mentioned STS results (Fig. 4b).

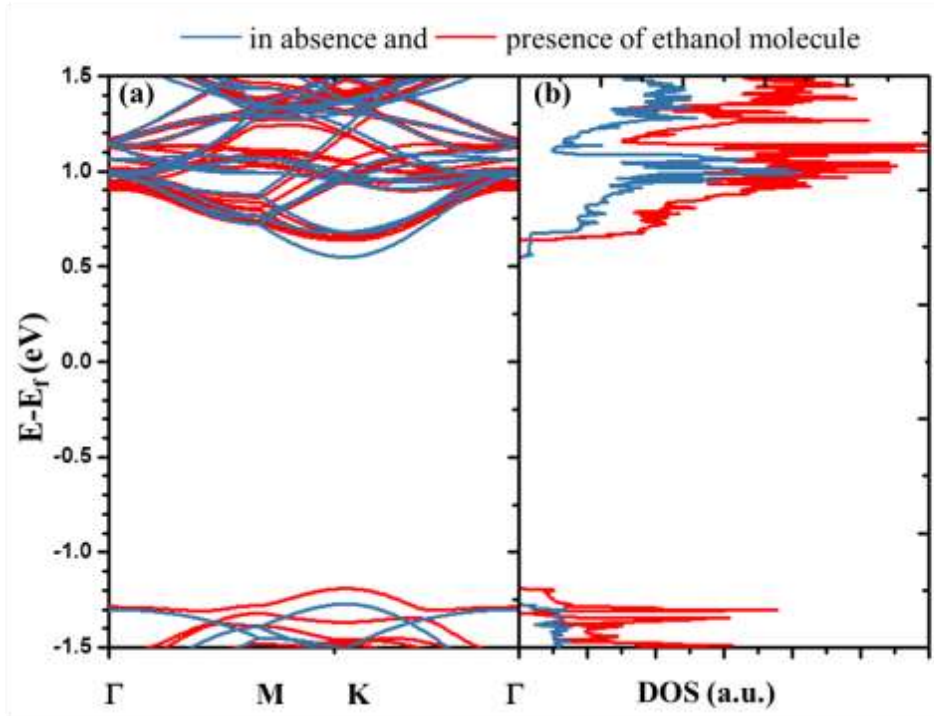

**Figure5:** Simulated band structure (a) and LDOS (b) of MoS₂ monolayer in presence (red) and absence (blue) of ethanol molecule

Small shift of the valance band edge toward lower energy and increment in the DOS near the valence and conduction band edge are clearly visible. According to Figure 5 the most affected k-point patch is in Γ while the gap occurs at K point. For further investigation, electron density (Bloch states) on the edge of the conduction band is calculated at the K and Γ points as is demonstrated in Figure 6.

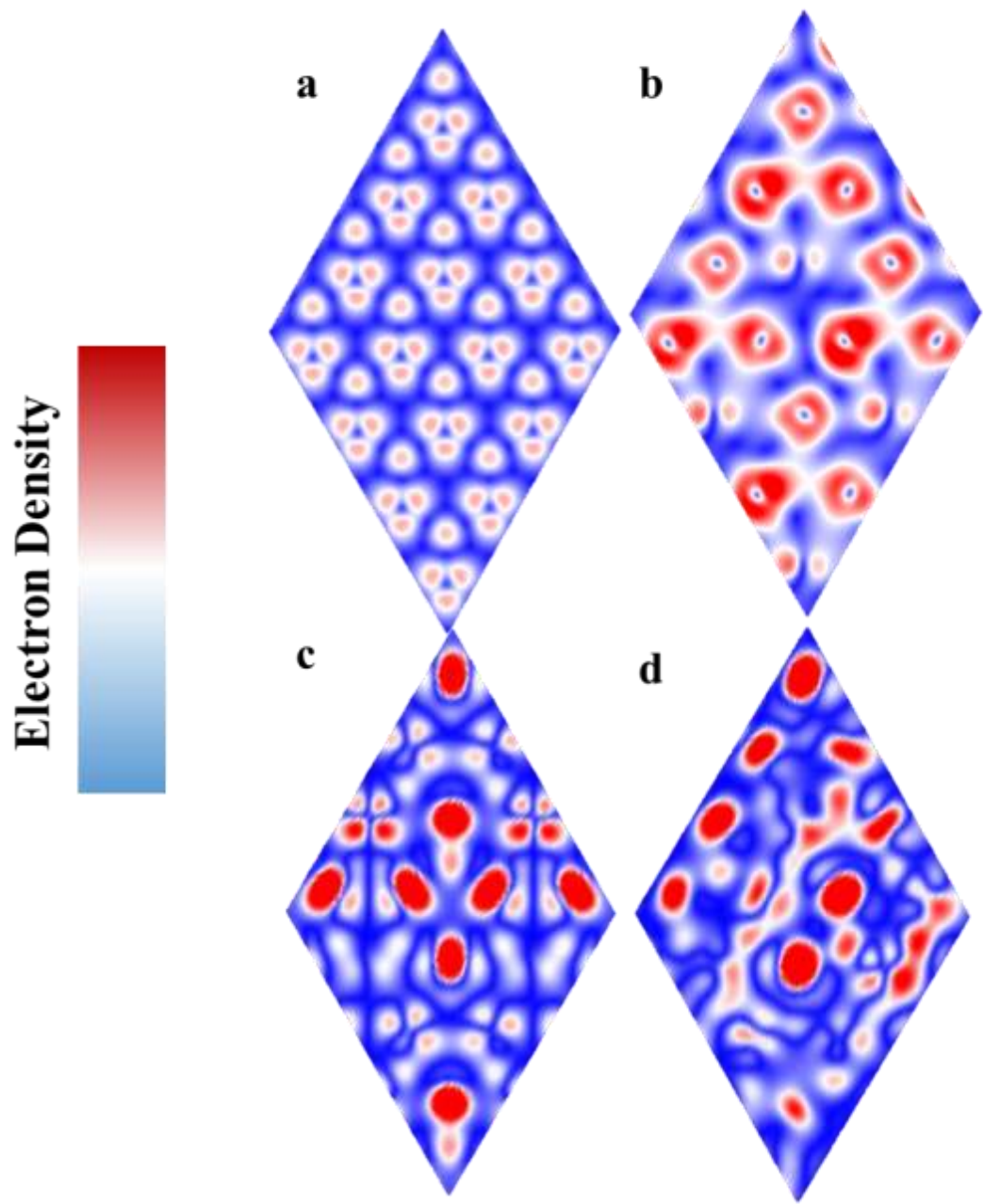

**Figure6:** Bloch states in edge of conduction band at K (a, b) and Γ points (c, d) in absence (a, c) and in presence (b, d) of one ethanol molecule.

$MoS_2$ monolayer shows completely uniform symmetric electron density at the edge of valence and conduction band in both points. In the presence of ethanol, the patterns are completely distorted. Proximity of ethanol molecules to the surface cause an electron density combination that leads to carrier transport being enhanced through sheets. This is the reason behind increment in conductivity due to ethanol exposure. Simulated current voltage characteristic of a resistance gas sensor based on a $MoS_2$ sheet indicate current increases i.e. carrier hopping through orbitals in presence of ethanol (Figure S3(b)).

**Conclusion**

In conclusion we synthesized nanosheets by liquid exfoliation. FTIR analyses have shown that the ethanol molecules physisorbed on the $MoS_2$ surface STS results have shown that ethanol molecules increase the DOS in edges of valance and conduction bands. On the other hand, the simulation results indicate that the main changes in the DOS occurred at the $\Gamma$ point. Investigation of orbital shapes at this point shows that orbital combination has occurred, which can lead to increase of carrier transport in $MoS_2$ monolayer in the presence of ethanol molecules. Our results showed STS is very useful method that may shed light on the mechanism behind gas surface interaction of 2D materials such as $MoS_2$.

**Acknowledgements**

The authors acknowledge the Iranian National Science Foundation (INSF) for support.

**Supporting Information.**

**I. Density functional theory**

To explore our results, simulation based on density functional theory has been applied on the system. A fully relaxed super-cell of MoS$_2$ containing 48 atoms has been considered as the base model for our investigation. In order to consider the effect of ethanol molecule, a fully relaxed pristine MoS$_2$ was added to the base model and fully relaxed till the force on each atom was less than 0.01eV. The Local-density approximations (LDA) with Perdew-Zunger (PZ) parametrization have been used to define the interactions in the system [1]. The Brillouin Zone samplings are considered using 40×40×1 Monkhorost-Pack grid for calculations and 80×80×1 grides for density of state (DOS). All calculation has been performed using Siesta Package [2]. To investigate the effect of ethanol on current-voltage characteristic of MoS$_2$, we also simulate the device containing 10 pristine unit cells of MoS$_2$ as channel and two unit cells for each electrodes act as drain and source as shown in Figure S3. The LDA-PZ has been considered again with $1\times10\times100$ k-point sampling along the channel. The Single zetta polarized has been used in our calculation.

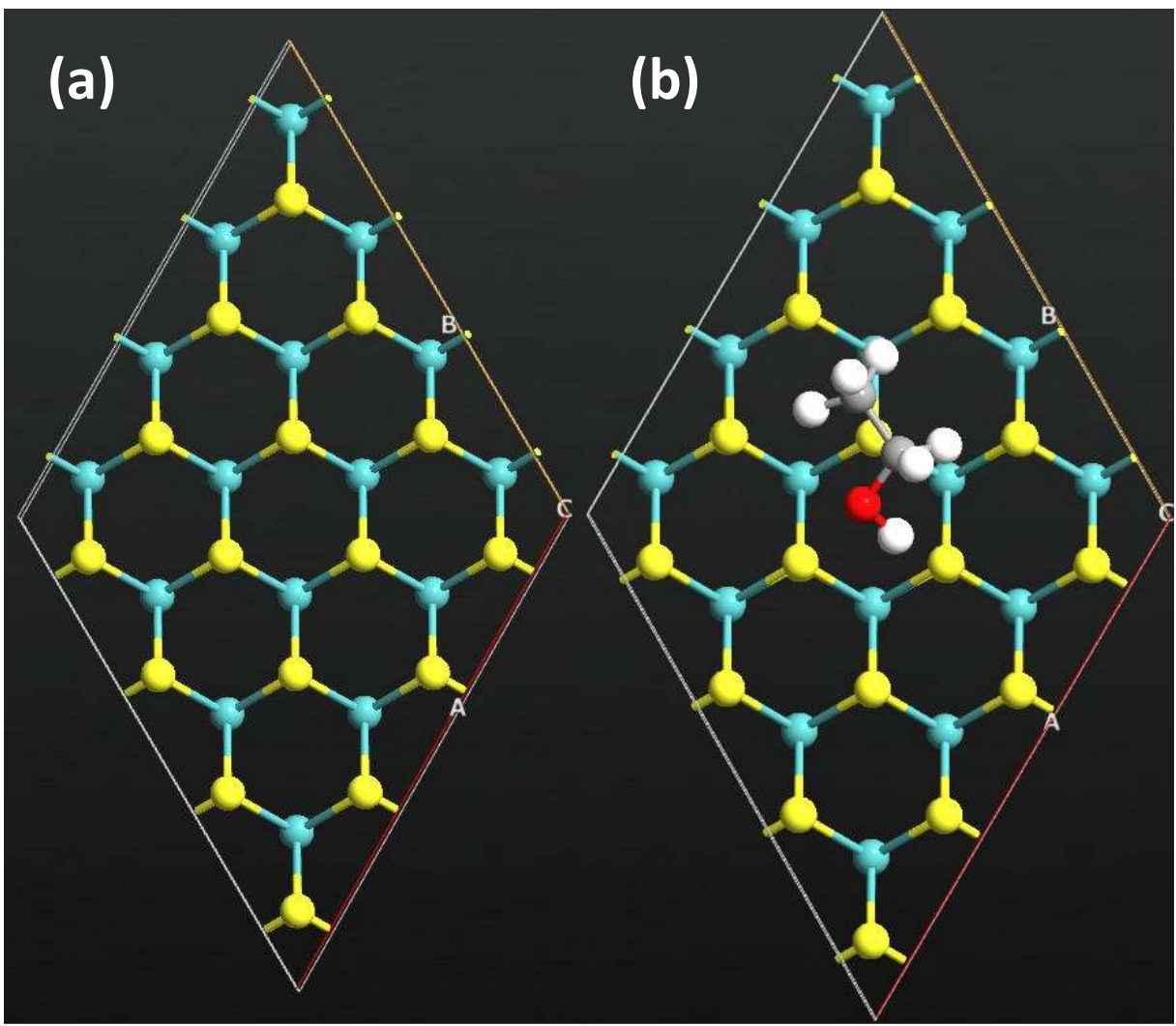

**Figure S1**: Optimized MoS$_2$ structure in absence (a) and presence (b) of ethanol.

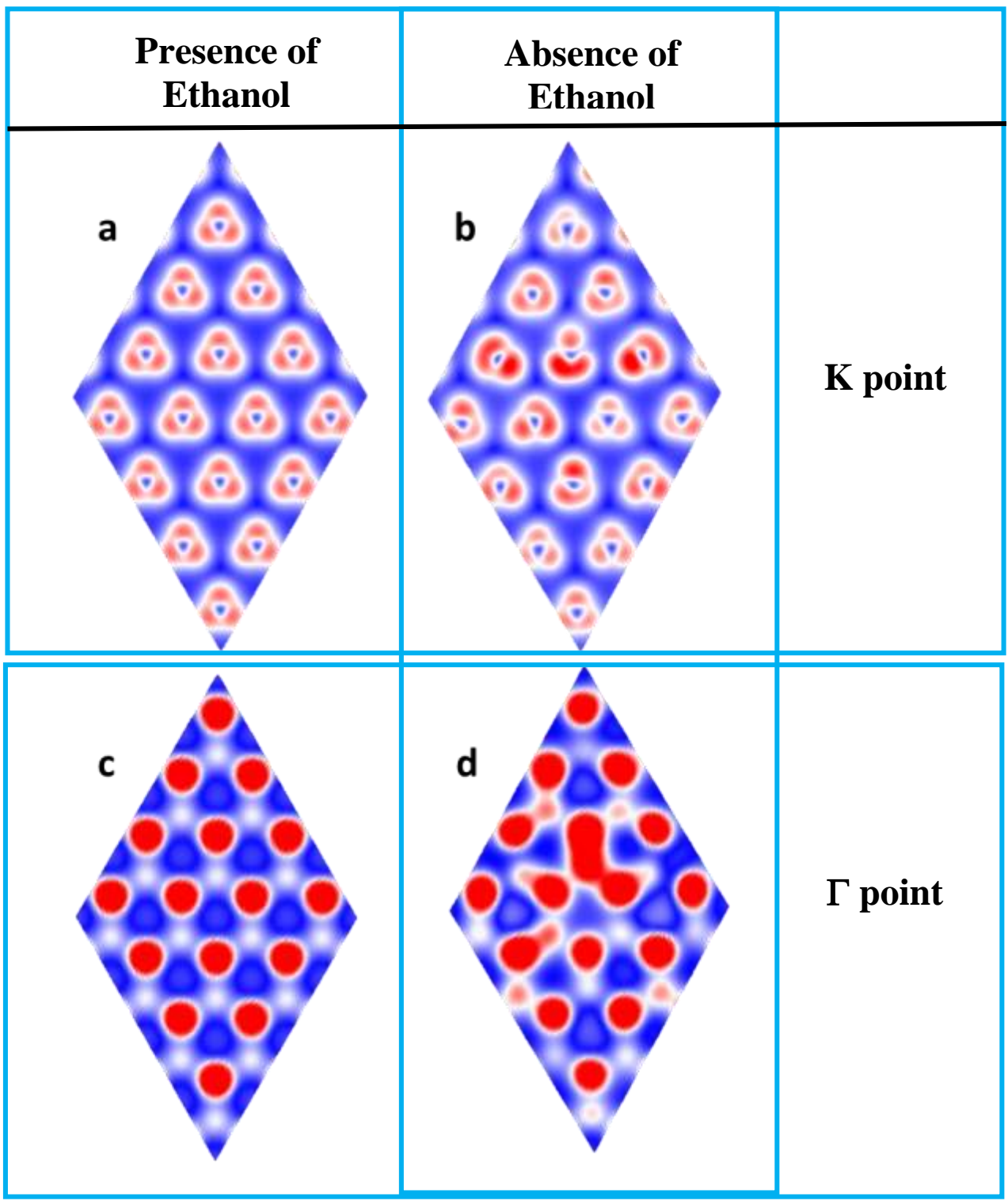

**Figure S2**: Bloch states in edge of valence band at K point (a, b) and gamma point (c, d) in presence (b, d) and absence (a, c) of one ethanol molecule at gamma point.

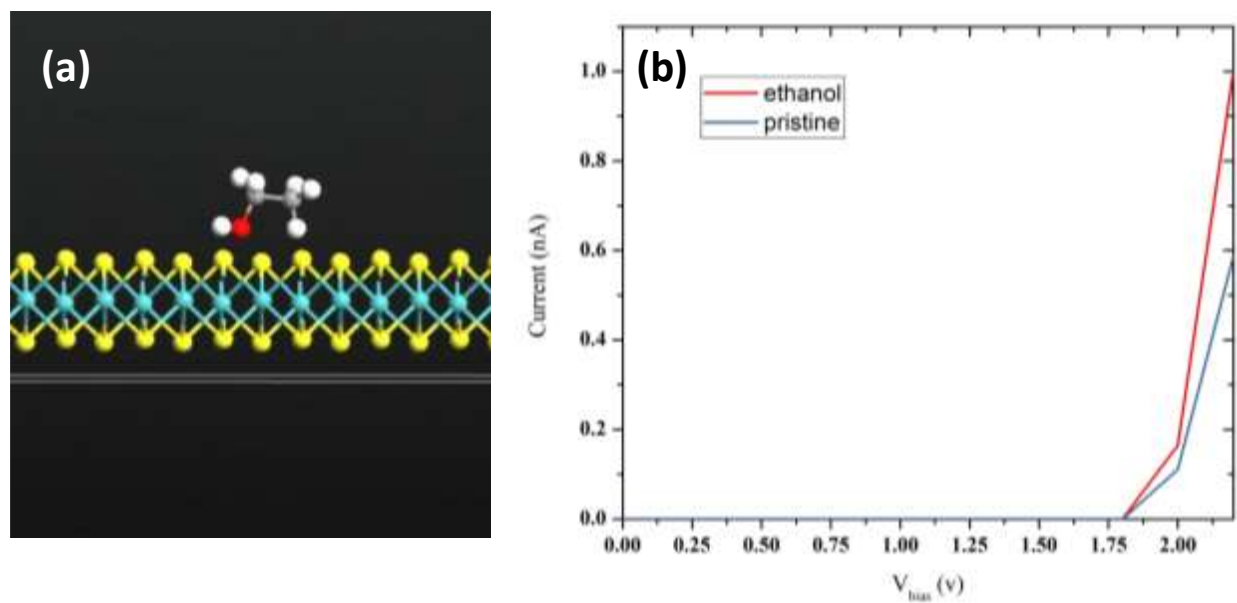

**Figure S3**: Simulated device structure (a) and I-V spectra (b).

The only point that remains is the mysterious small peaks near the edge of the valence band in Figure 3. These peaks have previously been reported for pristine $MoS_2$ and are related to the sulfur vacancy [3]. We simulate such condition and our results are in perfect agreement with previous reports [3]. The famous tip effect changes the state of s-vacancy defect which is existed near the conduction band (in simulation) to the valence band as discussed in detail before [3]. We consider the effect of s-vacancy on the adsorption of ethanol. Figure S4 shows the optimized structure and DOS. As can be seen, the s-vacancy is a desirable position for van der Waals adsorption of ethanol and no strong bond is created due to s-vacancy which agrees with our FTIR. The small alterations in the band gap still exist. The only visible change is the alteration of defect state to other energies. The number of subbands due to s-vacancy and the ethanol inside the bandgap are changed. There are two visible subbands near conduction band and a subband near the valence one. These bonds are changed due to the famous tip effect and shift to lower energies. The two subbands near the conduction can easily be distinguished in Figure 3 near the valence band. These changes due to vacancy alter the transmission and combination of orbitals but the discussion on the main effect of combination of orbitals is still valid and can be considered as the main reason for change in the conduction of MoS$_2$ in presence of ethanol.

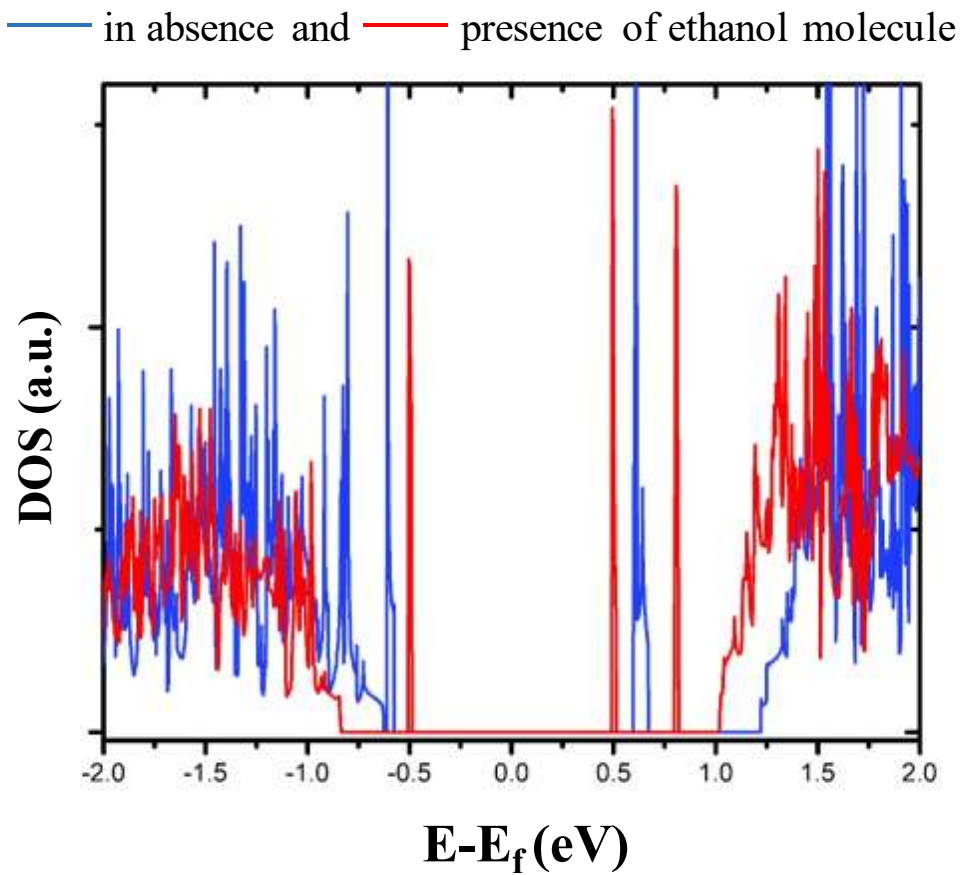

**Figure S4:** LDOS of MoS$_2$ with defect

**II. Exfoliation**

As mentioned in the article, MoS$_2$ powder exfoliated in NMP solvent by probe sonication. Figure S5 shows nanosheets at NMP after exfoliation.

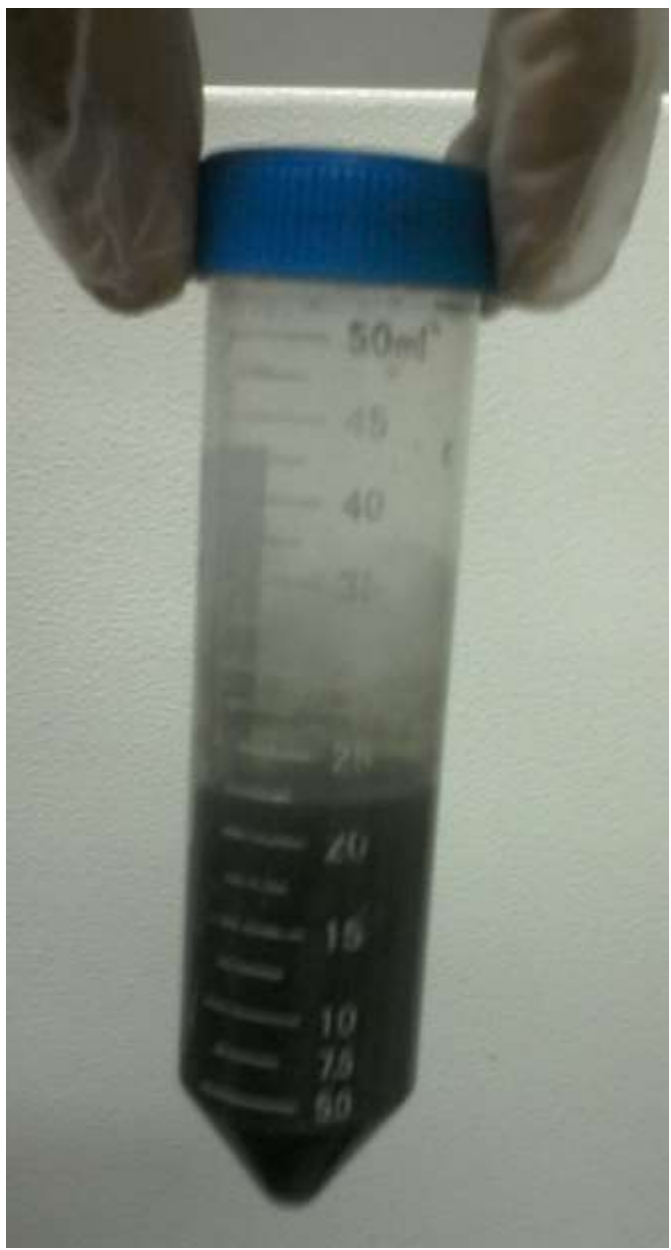

**Figure S5.** Exfoliated MoS2 nanosheets at NMP solvent.

**III. Scanning Tunneling Spectroscopy**

To calculate the DOS, the tunneling current is measured in terms of sample voltage. DOS are obtained using the following relationships.

$$DOS \propto \frac{dI/dV}{I/V} = \frac{\Delta I/\Delta V}{\overline{I}/\overline{V}} \qquad (1)$$

$$\Delta I = I_{n+1} - I_n \qquad (2)$$

$$\Delta V = V_{n+1} - V_n \qquad (3)$$

$$\overline{I} = \frac{I_{n+1} + I_n}{2} \qquad (4)$$

$$\overline{V} = \frac{V_{n+1} + V_n}{2} \qquad (5)$$

Figure S6 shows tunneling current and DOS in terms of voltage for MoS$_2$ monolayer.

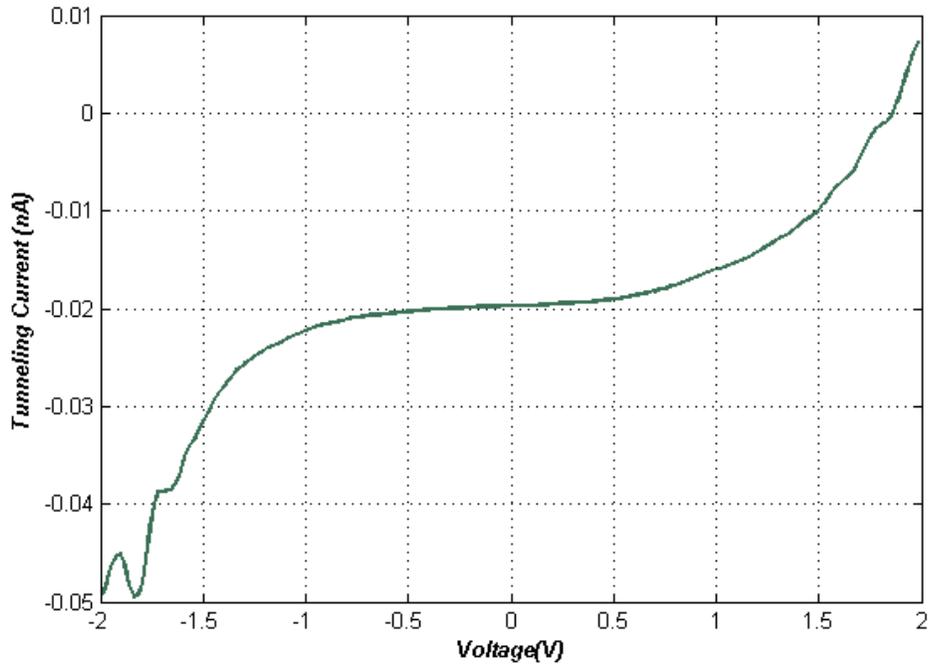

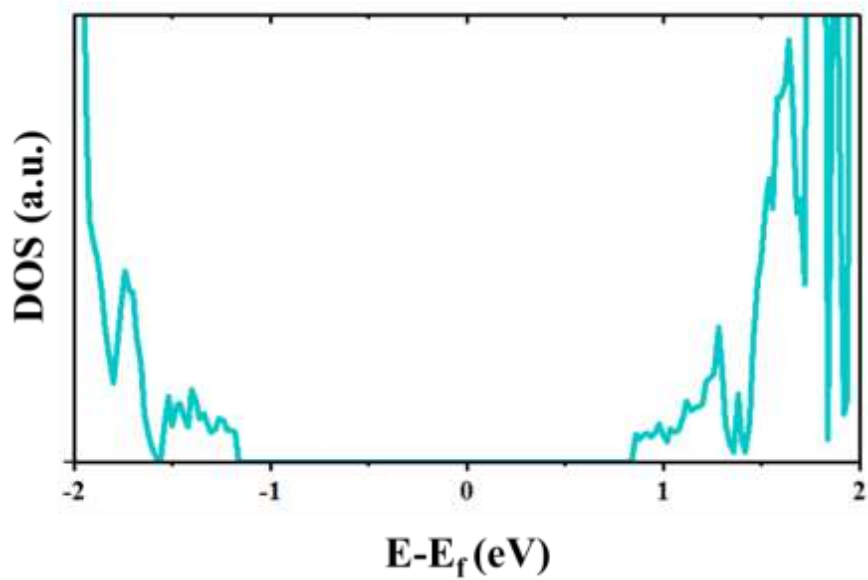

**Figure S6.** (a) tunneling current in terms of voltage and (b) DOS in terms of voltage for MoS$_2$ monolayer.

**Supporting reference**